\documentclass[12pt]{article}
\usepackage[dvips]{graphicx,color}
\usepackage{amsmath,bm,calc}
\usepackage[psamsfonts]{amssymb}
\begin{document}

\begin{titlepage}
\title{Hartree-Fock-Bogolyubov calculations
with Gaussian expansion method}

\author{H. Nakada\\
\textit{Department of Physics, Faculty of Science, Chiba University,}
\vspace*{-3mm}\\
\textit{Yayoi-cho 1-33, Inage, Chiba 263-8522, Japan}}

\date{\today}
\maketitle
\thispagestyle{empty}

\begin{abstract}
We extensively develop an algorithm
of implementing the Hartree-Fock-Bogolyubov calculations,
in which the Gaussian expansion method is employed.
This algorithm is advantageous in describing the energy-dependent
exponential and oscillatory asymptotics
of the quasiparticle wave functions at large $r$,
and in handling various effective interactions
including those with finite ranges.
We apply the present method to the oxygen isotopes
with the Gogny interaction, keeping the spherical symmetry.
In respect to the new magic numbers,
effects of the pair correlation on the $N=16$ and $32$ nuclei
are investigated.
\end{abstract}


\vspace*{3mm}\noindent
\end{titlepage}

\pagestyle{plain}

\section{Introduction}
\label{sec:intro}

Experimental data using the radioactive beams
have disclosed exotic natures of atomic nuclei
far from the $\beta$-stability.
Some of the data cast questions
on the conventional nuclear structure theories,
which have been founded mainly on top of the data
along the $\beta$-stability line.
Density profile of nuclei near the proton or neutron drip-line
is remarkably different from that of the stable nuclei,
sometimes making halos~\cite{ref:halo}.
Furthermore, new magic numbers, \textit{e.g.} $N=16$ and $32$,
have been reported in the region
off the $\beta$-stability line~\cite{ref:magic}.
A property of the effective interaction,
which has not been investigated sufficiently,
seems to play a role in the variation
of magic numbers~\cite{ref:Vst,ref:Nak03}.

The mean-field theories provide us with a good first approximation
for bound states of nuclei.
The Hartree-Fock (HF) theory,
through which we can construct the nuclear mean-field
in a self-consistent manner,
is a desirable tool to obtain the single-particle (s.p.) orbits of nuclei
from microscopic standpoints.
In addition, the HF theory is useful in investigating
basic and global natures of effective interactions.
However, the pair correlation between identical nucleons
yields a significant correction to the HF approximation.
The pair correlation, which gives rise to the superfluidity,
can be treated self-consistently
in the Hartree-Fock-Bogolyubov (HFB) theory~\cite{ref:HFB}.
For nuclei in the vicinity of the drip line,
appropriate treatment of the wave-function asymptotics at large $r$
may be important,
as has been shown in the HF framework~\cite{ref:NS02}.
If the pairing is taken into consideration,
the s.p. levels that are far below the Fermi energy in the HF scheme
couple to the continuum.
In applying the HFB approximation to nuclei near the drip line,
we need a method capable of handling the wave-function asymptotics
for bound nucleons and those for nucleons in the continuum simultaneously.
While this point has been argued
in association with the zero-range Skyrme interaction~\cite{ref:THO},
it has not been considered sufficiently for finite-range interactions,
primarily because no efficient methods
to treat the wave-function asymptotics have been known
in the case that the non-local nuclear currents are required.

We recently proposed a new computational method~\cite{ref:NS02}
to implement the HF calculations,
by extensively applying the Gaussian expansion method
(GEM)~\cite{ref:Kam88}.
It was demonstrated that the energy-dependent exponential asymptotics
of wave functions are handled efficiently,
even with finite-range interactions.
In this article we further extend the method to the HFB calculations.
Both the exponential and the oscillatory asymptotics
can be practically represented in a single set of Gaussian bases.
This method is easily adaptable to calculations
with finite-range interactions,
and therefore will be useful in studying effective interactions
to describe nuclear properties including the pair correlation.
The present method is applied to the oxygen isotopes in practice,
keeping the spherical symmetry.
The interaction-dependence of the shell structure
near $N=16$ and $32$ is also investigated,
in the light of the pair correlation.

\section{Single-particle bases and HFB equation}
\label{sec:basis}

In this paper we assume the mean fields to be spherically symmetric
and to preserve the parity.
Though almost straightforward,
extension to the deformed cases will be left as a future study.

To represent the s.p. wave functions,
we introduce bases having the following form:
\begin{eqnarray} \varphi_{\nu\ell jm}({\mathbf r})
&=& R_{\nu\ell j}(r)\,[Y^{(\ell)}(\hat{\mathbf r})
\chi_\sigma]^{(j)}_m\,; \nonumber\\
R_{\nu\ell j}(r) &=& \mathcal{N}_{\nu\ell j}\,r^\ell\exp(-\nu r^2)\,.
\label{eq:basis} \end{eqnarray}
Here $Y^{(\ell)}(\hat{\mathbf r})$ expresses the spherical harmonics
and $\chi_\sigma$ the spin wave function.
We drop the isospin index without confusion.
The index $\nu$ corresponds to the range parameter
of the Gaussian basis.
We allow the range parameter $\nu$ to be complex~\cite{ref:GEM},
and hereafter denote $\textrm{Re}(\nu)$ and $\textrm{Im}(\nu)$
by $\nu_\mathrm{r}$ and $\nu_\mathrm{i}$, respectively.
Note that $\nu_\mathrm{r}>0$.
The constant $\mathcal{N}_{\nu\ell j}$ is determined by
\begin{equation} \mathcal{N}_{\nu\ell j}
= \frac{2^{\ell+\frac{7}{4}}}{\pi^\frac{1}{4}\sqrt{(2\ell+1)!!}}\,
 \,\nu_\mathrm{r}^\frac{2\ell+3}{4},
\end{equation}
so as for $\langle\varphi_{\nu\ell jm}|\varphi_{\nu\ell jm}
\rangle$ to be unity.
The bases of Eq.~(\ref{eq:basis}) with different $\nu$'s
are not orthogonal to one another.
The norm matrix for each $(\ell,j)$ is composed of the elements of
\begin{equation} N^{(\ell j)}_{\nu\nu'}
= \langle\varphi_{\nu\ell jm}|\varphi_{\nu'\ell jm}\rangle
= \left(\frac{2\sqrt{\nu_\mathrm{r}\nu'_\mathrm{r}}}
   {\nu^\ast+\nu'}\right)^{\ell+\frac{3}{2}}\,. 
\label{eq:norm}\end{equation}
In the GEM we take $\nu$'s belonging to a geometric progression.
The GEM basis-set with real $\nu$'s has been shown to work efficiently
in the HF calculations~\cite{ref:NS02}
as well as in solving few-body problems~\cite{ref:KKF},
including the cases of loosely bound systems.
The exponential decrease of density at large $r$
is described to a good approximation
by a superposition of the Gaussians with various ranges.
For a complex $\nu$, the linear combinations
of $R_{\nu\ell j}$ and $R_{\nu^\ast\ell j}$ clarify
the oscillating structure as
\begin{eqnarray} [R_{\nu\ell j}(r) + R_{\nu^\ast\ell j}(r)]/2
 &=& \mathcal{N}_{\nu\ell j}\,r^\ell\exp(-\nu_\mathrm{r}r^2)\,
  \cos(\nu_\mathrm{i}r^2)\,,\nonumber\\
 ~[R_{\nu\ell j}(r) - R_{\nu^\ast\ell j}(r)]/2i
 &=& \mathcal{N}_{\nu\ell j}\,r^\ell\exp(-\nu_\mathrm{r}r^2)\,
  \sin(\nu_\mathrm{i}r^2)\,.\label{eq:basis2}
\end{eqnarray}
Indeed, by superposing the complex-range Gaussian bases,
oscillating behavior of the wave functions in the scattering states
as well as in the states with high-nodal wave functions
can be expressed efficiently~\cite{ref:GEM}.

To formulate the HFB theory for non-orthogonal bases,
we consider the creation (annihilation) operator
$c_{\nu\ell jm}^\dagger$ ($c_{\nu\ell jm}$),
which is associated with $\varphi_{\nu\ell jm}$
and obeys the non-canonical commutation relations,
\begin{equation} \{c_{\nu\ell jm}, c_{\nu'\ell'j'm'}^\dagger\}
 = \delta_{\ell\ell'} \delta_{jj'} \delta_{mm'} N^{(\ell j)}_{\nu\nu'}
  \,,\quad
 \{c_{\nu\ell jm}, c_{\nu'\ell'j'm'}\} =
 \{c_{\nu\ell jm}^\dagger, c_{\nu'\ell'j'm'}^\dagger\} = 0\,.
\end{equation}
The HFB equation for non-orthogonal bases is derived generally
in Appendix~\ref{app:HFBeq},
and it is reduced to the equation with the $\ell$ and $j$ conservation
in Appendix~\ref{app:sphHFB}.
Under the spherical symmetry,
which implies that $\ell$ and $j$ as well as $m$ are conserved,
the generalized Bogolyubov transformation is written as
\begin{equation}
 \alpha_{n\ell jm}^\dagger = \sum_\nu \left[
  U^{(\ell j)}_{\nu n} c_{\nu\ell jm}^\dagger
  + V^{(\ell j)}_{\nu n} c_{\nu\overline{\ell jm}} \right]\,,
\end{equation}
where the s.p. state $(\nu\overline{\ell jm})$ stands for
the time reversal to $(\nu\ell jm)$.
The solution of the HFB equation (\ref{HFBeq2})
gives the quasiparticle (q.p.) state represented by $(n\ell jm)$,
whose eigenvalue, $\epsilon_{n\ell j}$, is called quasiparticle energy.
In $Z$ or $N=\mathrm{odd}$ nuclei,
we should replace $(\mathsf{U}^{(\ell j)},-\mathsf{V}^{(\ell j)})$
by $(\mathsf{V}^{(\ell j)\ast},\mathsf{U}^{(\ell j)\ast})$
in Eq.~(\ref{Rmat2}) for a nucleon
which has the lowest $\epsilon_{n\ell j}$.
With the spherical symmetry,
this implies addition of
$\frac{1}{2j+1}(U^{(\ell j)}_{\nu n} U^{(\ell j)^\ast}_{\nu' n}
-V^{(\ell j)\ast}_{\nu n} V^{(\ell j)}_{\nu' n})$
to $\rho^{(\ell j)}_{\nu\nu'}$
and $-\frac{1}{2j+1}(U^{(\ell j)}_{\nu n} V^{(\ell j)^\ast}_{\nu' n}
+V^{(\ell j)\ast}_{\nu n} U^{(\ell j)}_{\nu' n})$
to $\kappa^{(\ell j)}_{\nu\nu'}$ in Eq.~(\ref{Rmat2}).
With this modification, the form of the HFB equation (\ref{HFBeq2})
does not change.
If we use the linear combination of Eq.~(\ref{eq:basis2})
instead of the complex-range Gaussian bases,
all matrices in Eq.~(\ref{HFBeq2}) can be taken to be real.

\section{Asymptotic behavior of quasiparticle wave functions}
\label{sec:asymp}

We next consider the asymptotic behavior of the q.p. wave functions
at large $r$.
The HFB equation (\ref{HFBeq2}) is easily represented
in terms of the radial coordinate $r$,
by converting the label of the basis $\nu$ to $r$.
For the time being we consider the HFB equation for neutrons.
At sufficiently large $r$, the nuclear force becomes negligible,
and the following asymptotic equations are derived~\cite{ref:SkP},
\begin{eqnarray}
 \left(-\frac{1}{2M}\frac{d^2}{dr^2} - \lambda\right)\,
  [r\,U^{(\ell j)}_n(r)]
 &\approx& \epsilon_{n\ell j} [r\,U^{(\ell j)}_n(r)]\,,\nonumber\\
 \left(-\frac{1}{2M}\frac{d^2}{dr^2} - \lambda\right)\,
  [r\,V^{(\ell j)}_n(r)]
 &\approx& -\epsilon_{n\ell j} [r\,V^{(\ell j)}_n(r)]\,.
\label{HFBeq-asymp}\end{eqnarray}
As long as the nucleus is bound,
the chemical potential $\lambda$ should be negative.
Since we take $\epsilon_{n\ell j}$ to be positive,
Eq.~(\ref{HFBeq-asymp}) derives the asymptotic form,
\begin{equation}
 r\,V^{(\ell j)}_n(r)\approx \exp(-\eta^{(\ell j)}_{n+} r)\,,\quad
 r\,U^{(\ell j)}_n(r)\approx\left\{\begin{array}{ccc}
 \exp(-\eta^{(\ell j)}_{n-} r) &
  \mbox{for}& \lambda+\epsilon_{n\ell j} < 0 \\
 \cos(p^{(\ell j)}_{n} r + \theta^{(\ell j)}_{n}) &
  \mbox{for}& \lambda+\epsilon_{n\ell j} > 0 \end{array}\right.,
\label{qpwf-asymp}\end{equation}
where
$\eta^{(\ell j)}_{n\pm} = \sqrt{2M(-\lambda\pm\epsilon_{n\ell j})}$,
$p^{(\ell j)}_{n} = \sqrt{2M(\lambda+\epsilon_{n\ell j})}$
and $\theta^{(\ell j)}_{n}$ is an appropriate real number.
Note that $\eta^{(\ell j)}_{n+} > \eta^{(\ell j)}_{n-}$.
Corresponding to the asymptotic behavior of $U^{(\ell j)}_n(r)$,
the q.p. energies are discrete
only for $0\leq\epsilon_{n\ell j}<-\lambda$.
Thus the deeply bound s.p. states in the HF approximation
are embedded in the continuum in the HFB theory.
It is now obvious that, in the HFB calculations,
we have to treat both the exponential and the oscillatory asymptotics
which depend on the q.p. energy.
In Ref.~\cite{ref:THO}, these asymptotics are handled
by adopting the transformed harmonic oscillator (THO) bases.
However, it is not easy to apply the THO bases
if the non-local currents are required,
as in the case of the finite-range interactions.
As will be demonstrated in Sec.~\ref{sec:test},
the GEM provides a practical method
in handling the energy-dependent asymptotics
even in the presence of the non-local currents.

The asymptotic behavior of the neutron density
and pair current~\cite{ref:BDP00} is derived
from Eq.~(\ref{qpwf-asymp}).
The density matrix and the pairing tensor in Eq.~(\ref{Rmat2})
are converted to the $r$-representation as
\begin{eqnarray} \rho^{(\ell j)} (r,r')
 &=& \sum_n V^{(\ell j)\ast}_n (r)\,V^{(\ell j)}_n (r')\,,\nonumber\\
 \kappa^{(\ell j)} (r,r')
 &=& \sum_n V^{(\ell j)\ast}_n(r)\,U^{(\ell j)}_n(r')\,,
\label{Rmat3}\end{eqnarray}
for $N=\mathrm{even}$ nuclei.
The density and the pair current are then defined by
\begin{eqnarray}
 \rho(r) &=& \sum_{\ell j} (2j+1)\,\rho^{(\ell j)}(r,r)\,,\nonumber\\
 \kappa(r) &=& \sum_{\ell j} (2j+1)\,\kappa^{(\ell j)}(r,r)\,.
\label{current}\end{eqnarray}
Denoting the smallest q.p. energy by $\epsilon_\mathrm{min}$,
we immediately obtain the asymptotic form of $\rho(r)$ as
\begin{equation}
 r^2\rho(r)\approx \exp(-2\eta^\mathrm{min}_+ r)\,,
\label{dns-asymp}\end{equation}
where $\eta^\mathrm{min}_+ = \sqrt{2M(-\lambda+\epsilon_\mathrm{min})}$.
In order for an $N=\mathrm{odd}$ nucleus to be bound,
the condition $\lambda+\epsilon_\mathrm{min}<0$ must be fulfilled,
because $V^{(\ell j)}_n (r)$ is replaced by $U^{(\ell j)}_n (r)$
for the last neutron.
The contribution of the last neutron yields the asymptotic form
\begin{equation}
 r^2\rho(r)\approx \exp(-2\eta^\mathrm{min}_- r)\,,
\label{dns-asymp2}\end{equation}
which damps more slowly than the form in Eq.~(\ref{dns-asymp}).

As pointed out in Ref.~\cite{ref:BDP00},
a complication may arise in the close vicinity of the drip line.
If $-\lambda$ is almost vanishing and
no q.p. level satisfies $\lambda+\epsilon_{n\ell j}<0$,
it is difficult to draw definite conclusion
on the asymptotic behavior of $\rho(r)$.
For $N=\mathrm{even}$ nuclei,
arguments based on the q.p. energy suggest
that the component of $\lambda+\epsilon_{n\ell j}\rightarrow +0$
becomes dominant at the $r\rightarrow\infty$ limit,
giving the asymptotics of
$r^2\,\rho(r)\approx \exp\!\left[-2\sqrt{4M(-\lambda)}\,r\right]$.
However, the spectroscopic amplitude of this component
may be negligibly small in the physically interesting region.
The actual situation is in the balance
between the damping factor and the spectroscopic amplitude
of the q.p. levels near $-\lambda$.
Moreover, because of this subtlety,
in practical calculations the asymptotic behavior of $\rho(r)$
could be sensitive to the details of the method,
even if individual q.p. wave function has the asymptotics
consistent with its q.p. energy.
Even with the high adaptability of the GEM
to the wave-function asymptotics at large $r$,
it is beyond scope of the present study
to obtain reliable asymptotics of $\rho(r)$
for nuclei with $\lambda+\epsilon_\mathrm{min}>0$.
Similar arguments apply to $N=\mathrm{odd}$ nuclei.

The asymptotics for $\kappa(r)$ could also be complicated.
We separately consider contribution of the discrete q.p. states
and that of the q.p. continuum.
The discrete q.p. states, which satisfy $\lambda+\epsilon_{n\ell j}<0$,
contribute to $\kappa(r)$ asymptotically with the damping factor
$\exp\!\left[-(\eta^{(\ell j)}_{n+} + \eta^{(\ell j)}_{n-}) r\right]$.
Because $\eta^{(\ell j)}_{n+} + \eta^{(\ell j)}_{n-}$
is a decreasing function of $\epsilon_{n\ell j}$ for a fixed $\lambda$,
the q.p. level just below $-\lambda$
dominates over the other discrete states at large $r$.
The damping factor becomes $\exp\!\left[-2\sqrt{4M(-\lambda)}\,r\right]$
at the $\lambda+\epsilon_{n\ell j}\rightarrow -0$ limit.
In the states having $\lambda+\epsilon_{n\ell j}>0$,
we have a damping factor $\exp(-\eta^{(\ell j)}_{n+} r)$,
which is multiplied by the oscillating factor
$\cos(p^{(\ell j)}_{n} r + \theta^{(\ell j)}_{n})$, at large $r$.
Hence the q.p. level closest to $-\lambda$
has dominant contribution to $\kappa(r)$
at the $r\rightarrow\infty$ limit.
The component of $\lambda+\epsilon_{n\ell j}\rightarrow +0$ gives
the most slowly damping factor
$\exp\!\left[-\sqrt{4M(-\lambda)}\,r\right]$.
However, similarly to $\rho(r)$ in the $\lambda\approx 0$ case,
spectroscopic amplitudes of the q.p. levels become relevant.
This complicates description of the asymptotic behavior
of $\kappa(r)$ near the neutron drip line.

For proton wave functions,
the Coulomb interaction influences the wave functions at large $r$,
although it does not affect the criterion
whether or not individual q.p. levels are discrete.
The above arguments for the neutron functions
can be applied if the asymptotic forms are properly modified.
Nevertheless, the complication near the drip line
is expected to be less serious than for neutron wave functions,
owing to the presence of the Coulomb barrier.

\section{Effective interaction}
\label{sec:effint}

We shall consider the effective Hamiltonian
comprised of the kinetic energy and the effective two-body interaction,
\begin{equation}
\hat{H} = \hat{K} + \hat{V}\,;\quad
 \hat{K} = \sum_i \frac{\mathbf{p}_i^2}{2M}\,,\quad
\hat{V} = \sum_{i<j} v_{ij}\,.
\label{eq:Hamil}\end{equation}
Here $i$ and $j$ are the indices of each nucleon.
The s.p. matrix element of the kinetic term is calculated as
\begin{equation} \langle \varphi_{\nu\ell jm}|
\frac{\mathbf{p}^2}{2M}|\varphi_{\nu'\ell jm}\rangle
= \frac{2\ell+3}{2M} \cdot \frac{2\nu^\ast \nu'} {\nu^\ast + \nu'}\,
 N^{(\ell j)}_{\nu\nu'}\,.
\label{eq:kin-me}\end{equation}
As in Refs.~\cite{ref:Nak03,ref:NS02},
the effective interaction $v_{ij}$ is assumed to have the following form,
\begin{eqnarray} v_{12} &=& v_{12}^{(\mathrm{C})}
 + v_{12}^{(\mathrm{LS})} + v_{12}^{(\mathrm{TN})}
 + v_{12}^{(\mathrm{DD})}\,;\nonumber\\
v_{12}^{(\mathrm{C})} &=& \sum_\mu (t_\mu^{(\mathrm{SE})} P_\mathrm{SE}
 + t_\mu^{(\mathrm{TE})} P_\mathrm{TE}
 + t_\mu^{(\mathrm{SO})} P_\mathrm{SO}
 + t_\mu^{(\mathrm{TO})} P_\mathrm{TO}) f_\mu^{(\mathrm{C})}(r_{12})\,,
 \nonumber\\
v_{12}^{(\mathrm{LS})} &=& \sum_\mu (t_\mu^{(\mathrm{LSE})} P_\mathrm{TE}
 + t_\mu^{(\mathrm{LSO})} P_\mathrm{TO}) f_\mu^{(\mathrm{LS})} (r_{12})\,
 \mathbf{L}_{12}\cdot(\mathbf{s}_1+\mathbf{s}_2)\,,\nonumber\\
v_{12}^{(\mathrm{TN})} &=& \sum_\mu (t_\mu^{(\mathrm{TNE})} P_\mathrm{TE}
 + t_\mu^{(\mathrm{TNO})} P_\mathrm{TO})
 f_\mu^{(\mathrm{TN})} (r_{12})\, r_{12}^2 S_{12}\,,\nonumber\\
v_{12}^{(\mathrm{DD})} &=& t_3 (1 + x_3 P_\sigma)
 [\rho(\mathbf{r}_1)]^\alpha \delta(\mathbf{r}_{12})\,,
\label{eq:effint}\end{eqnarray}
where $\mathbf{r}_{12}= \mathbf{r}_1 - \mathbf{r}_2$,
$r_{12}=|\mathbf{r}_{12}|$,
$\mathbf{p}_{12}= (\mathbf{p}_1 - \mathbf{p}_2)/2$,
$\mathbf{L}_{12}= \mathbf{r}_{12}\times\mathbf{p}_{12}$,
$\mathbf{s}_i$ is the nucleon spin operator,
and $S_{12}$ is the tensor operator
$S_{12}= 4\,[3(\mathbf{s}_1\cdot\hat{\mathbf{r}}_{12})
(\mathbf{s}_2\cdot\hat{\mathbf{r}}_{12})
- \mathbf{s}_1\cdot\mathbf{s}_2 ]$.
The projectors $P_\mathrm{SE}$, $P_\mathrm{TE}$, $P_\mathrm{SO}$
and $P_\mathrm{TO}$ are expressed as
\begin{eqnarray}
P_\mathrm{SE} = \frac{1-P_\sigma}{2}\,\frac{1+P_\tau}{2}\,,&&
P_\mathrm{TE} = \frac{1+P_\sigma}{2}\,\frac{1-P_\tau}{2}\,,\nonumber\\
P_\mathrm{SO} = \frac{1-P_\sigma}{2}\,\frac{1-P_\tau}{2}\,,&&
P_\mathrm{TO} = \frac{1+P_\sigma}{2}\,\frac{1+P_\tau}{2}\,,
\label{eq:proj_eo}\end{eqnarray}
in terms of the spin and isospin exchange operators
$P_\sigma$, $P_\tau$.
In Eq.~(\ref{eq:effint}),
$f_\mu$ represents an appropriate function,
$\mu$ stands for the parameter attached to the function,
and $t_\mu$ the coefficient.
By taking various functions for $f_\mu$,
Eq.~(\ref{eq:effint}) covers a wide variety of effective interactions.
The Gogny interaction~\cite{ref:Gogny} is obtained by setting
$f_\mu^{(\mathrm{C})}(r_{12}) = \exp(-\mu r_{12}^2)$,
$f^{(\mathrm{LS})} (r_{12})=\nabla^2\delta(\mathbf{r}_{12})$
and $v_{12}^{(\mathrm{TN})}=0$.
In Ref.~\cite{ref:Nak03} we have developed M3Y-type interactions,
in which we take $f_\mu^{(\mathrm{C,LS,TN})}(r_{12})
= \exp(-\mu r_{12})/\mu r_{12}$.

The matrix elements of $v_{12}^{(\mathrm{C,LS,TN})}$ can be calculated
by utilizing the Fourier transformation of $f_\mu(r_{12})$~\cite{ref:NS02}.
Most formulae in Ref.~\cite{ref:NS02} are straightforwardly applicable,
although the complex-range Gaussian bases
lead to a slight modification in the radial integral
(Eqs.~(24--26) in Ref.~\cite{ref:NS02}):
\begin{eqnarray}
\int_0^\infty r^2dr j_\lambda(kr)
 R_{\nu^\ast\ell j}(r) R_{\nu'\ell' j'}(r)
 \hspace*{8cm}\nonumber\\
 = \frac{\displaystyle 2^\frac{\ell+\ell'}{2}\,
  \left(\frac{\ell+\ell'-\lambda}{2}\right)!}
  {\sqrt{(2\ell+1)!! (2\ell'+1)!!}}\,
  \frac{\sqrt{2\nu_\mathrm{r}}^{\ell+\frac{3}{2}}
  \sqrt{2\nu'_\mathrm{r}}^{\ell'+\frac{3}{2}}}
  {(\nu^\ast+\nu')^\frac{\ell+\ell'+3}{2}}\cdot
  Q_{\nu^\ast\nu'}^\lambda\,
  L^{(\lambda+\frac{1}{2})}_\frac{\ell+\ell'-\lambda}{2}
  (Q_{\nu^\ast\nu'}^2)\,
  \exp(-Q_{\nu^\ast\nu'}^2) \,,
\label{eq:r-int-cent}\end{eqnarray}
where $L^{(\alpha)}_n(x)$ is the associated Laguerre polynomial and
\begin{equation}
 Q_{\nu^\ast\nu'} = \frac{k}{2(\nu^\ast+\nu')^\frac{1}{2}}\,.
\end{equation}
We need only the $\ell+\ell'+\lambda=\mathrm{even}$ cases
for the integral of Eq.~(\ref{eq:r-int-cent}).
In coding a computer program,
we prepare a subprogram for the $k$-integration
(see Eq.~(28) of Ref.~\cite{ref:NS02}),
which is the only part dependent on the form of $f_\mu(r_{12})$.
This allows us to handle various interaction forms
only by substituting the subprogram.
Although analytic formulae for the $k$-integration were given
in Ref.~\cite{ref:NS02},
their application sometimes causes serious round-off errors.
Numerical integration over $k$ is recommended to reduce the errors.

Mean-field approaches for nuclei are sometimes discussed
in connection to the density-functional theory (DFT).
Although the DFT is a rigorous theory,
we have to know a good density-functional form of energy
for practical applications.
The most popular density-functional form in nuclear structure physics
is based on the Skyrme interaction and its extensions,
whose parameters are adjusted to the known data.
Some of the Skyrme density functionals involve
currents that cannot be derived from two-body interactions.
Still the form of the Skyrme density functionals
is quite restricted,
since all currents are assumed to be local.
In order to know appropriate form of density functional for nuclei
covering large area of the nuclear chart,
it will be an important step to explore
finite-range effective two-body interactions.
It is noted that, whereas we consider two-body interactions
having the form of Eq.~(\ref{eq:effint}) in this paper,
the GEM algorithm is applicable also to the DFT,
which may not necessarily be connected to two-body interactions.

Once storing the interaction matrix elements
and having a trial set of $U^{(\ell j)}_{\nu n}$
and $V^{(\ell j)}_{\nu n}$,
we obtain the HFB equation (\ref{HFBeq2})
with $h^{(\ell j)}_{\nu\nu'}$ and $\Delta^{(\ell j)}_{\nu\nu'}$
defined in Eqs.~(\ref{spham0}) and (\ref{spham}),
and the equation is solved iteratively until convergence.

\section{Numerical examples}
\label{sec:test}

We now apply the present method of the HFB calculations to several nuclei,
assuming the spherical symmetry and the parity conservation.
The center-of-mass energy is fully removed before variation,
by subtracting both the one-body and two-body terms
from the Hamiltonian of Eq.~(\ref{eq:Hamil}).

\subsection{Application to oxygen isotopes
with Gogny interaction}

We first show results for oxygen isotopes,
using the standard D1S parameter-set~\cite{ref:D1S}
of the Gogny interaction.
While the real-range GEM basis-sets are efficient to describe
nodeless but broadly distributing wave functions,
the complex-range GEM sets seem suitable
for reproducing the oscillating behavior
required in reaction studies~\cite{ref:GEM}.
In the HFB calculations both types of wave functions
come into the problem.
We try the following three basis-sets,
a set comprised of real-range bases,
a set of complex-range bases and a set of their mixture;
\begin{displaymath}\begin{array}{lllll}
\mbox{Set~A\,:}& \nu_\mathrm{r}=\nu_\omega\,b^{-2\alpha}\,,&
b=1.20\,,&\, \nu_\mathrm{i}=0\,,& (\alpha=-2,-1,0,\cdots,K-3)\,;\\
\mbox{Set~B\,:}& \nu_\mathrm{r}=\nu_\omega\,b^{-2\alpha}\,,&
b=1.15\,,&\, {\displaystyle\frac{\nu_\mathrm{i}}{\nu_\mathrm{r}}
=\pm\frac{\pi}{2}}\cdot 0.6\,,& (\alpha=0,1,\cdots,K/2-1)\,;
\vspace*{2mm}\\
\mbox{Set~C\,:}& \nu_\mathrm{r}=\nu_\omega\,b^{-2\alpha}\,,&
b=1.25\,,&\!\! \left\{\begin{array}{l}\nu_\mathrm{i}=0\,,\\
 {\displaystyle\frac{\nu_\mathrm{i}}{\nu_\mathrm{r}}
=\pm\frac{\pi}{2}}\end{array}\right.&
\!\!\begin{array}{l}(\alpha=0,\cdots,K/2-1)\\
 (\alpha=0,\cdots,K/4-1)\end{array}\,,\\
\end{array}\end{displaymath}
where $\nu_\omega = M\omega/2\hbar$
with $\hbar\omega=41.2\times 24^{-1/3}\,{\rm MeV}$.
The parameter $K$ stands for number of bases for each $(\ell,j)$,
and is fixed to be $12$ for all the three sets.
We truncate the space, taking the $\ell\leq 4$ s.p. bases.
If the $\ell=5$ bases are added,
we have energy gain of $\lesssim 0.06\,\mathrm{MeV}$.

In practical calculations,
the common ratio $b$ of the GEM bases is restricted
so that the norm matrix should not have a singularity.
At a certain value of $b$ which is close to unity,
convergence in the iterative process becomes extremely slow
in numerical calculations with a given precision.
We here call this phenomenon `quasi-instability'.
If $b$ further approaches unity,
true instability comes out,
in which round-off errors lead to unphysical results,
because the norm matrix becomes nearly singular.
Presently the parameter $b$ is chosen for each set
so as to optimize the energy of $^{26}$O,
with avoiding the quasi-instability
in calculations using the double precision.
In the present calculation $^{26}$O is bound
although it lies beside the neutron drip line,
whereas the actual $^{26}$O seems unbound~\cite{ref:O26}.
By contrast, $^{28}$O is unbound
in the respect that the neutron chemical potential becomes positive,
which leads to the nuclear wave function distributing
up to the infinite distance.
For Set~B, the value of $\nu_\mathrm{i}/\nu_\mathrm{r}$
has also been tuned.
In the GEM algorithm of the mean-field calculations,
the parameters for the basis-set do not need to depend strongly
on mass number.
Using the same basis-set, we can obtain results
for all nuclides under consideration in a single run,
without recalculating the matrix elements of the effective interaction.

If the D1S force is applied to the pure neutron matter,
the energy per nucleon diverges with the negative sign
at the high density limit.
Due to this nature,
the true energy minimum in a finite nucleus lies at the configuration
in which all neutrons gather in the vicinity of the origin
({\it i.e.} the center-of-mass)
without overlap of the proton distribution.
To avoid the tunneling to this unphysical configuration,
the range of the GEM bases ($\nu_\mathrm{r}^{-1/2}$)
should not be too small.
All the present sets are chosen to meet this criterion.

We tabulate total energies of $^{14-26}$O
obtained from the HFB calculations using the three basis-sets,
in Table~\ref{tab:eng_O}.
Energies in the HF approximation are also presented.
The result for $^{14}$O of the HFB calculation with Set~A
is not shown, because the quasi-instability takes place.
The proton $0s$- and $0p$-shells are fully occupied,
forming the $Z=8$ closed core, in any results of Table~\ref{tab:eng_O}.
In $^{15-17,21,23-25}$O, neutrons lose the superfluidity,
irrespectively of the basis-sets.
Because the variational principle is satisfied in the HF and HFB theories,
the lower total energy indicates the better result for each nucleus.
The energies obtained from Sets~A and B are comparable,
with difference less than 10\,keV in most oxygen isotopes.
To be more precise,
while it depends on the nuclide which of Sets~A and B
gives lower energy in the HF calculation,
Set~B yields steadily lower energy than Set~A in the HFB calculation,
with the exception of $^{21}$O.
This is obviously because the energy gain due to the pairing
is larger in Set~B than in Set~A.
Giving energies lower than Sets~A and B by several tens keV,
Set~C is advantageous over the other sets
for any oxygen isotopes in both the HF and the HFB calculations.
Because the energy difference among the basis-sets
is contributed by individual q.p. or s.p. orbits,
difference in total energy will roughly be
proportional to the mass number,
growing significantly for heavier nuclei~\cite{ref:Shi_pv}.

\begin{figure}
\centerline{\includegraphics[width=15cm]{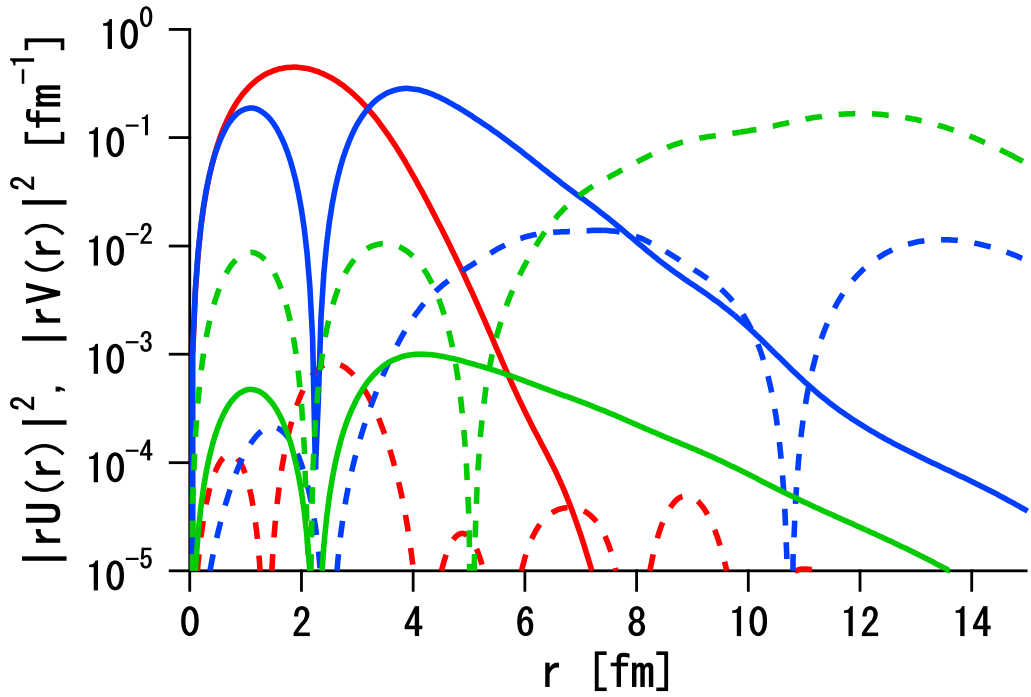}}
\vspace{2mm}
\caption{Wave functions of neutron $s_{1/2}$ q.p. levels
 in $^{26}$O calculated with the D1S interaction.
 The colors distinguish the q.p. levels:
 The red, blue and green lines show the wave functions of the levels
 corresponding to the $0s_{1/2}$, $1s_{1/2}$ and $2s_{1/2}$ orbits,
 respectively.
 The full (dashed) curves represent $|r\,V^{(\ell j)}_n(r)|^2$
 ($|r\,U^{(\ell j)}_n(r)|^2$) for each level.
\label{fig:qpwf}}
\end{figure}

\begin{figure}
\centerline{\includegraphics[width=13cm]{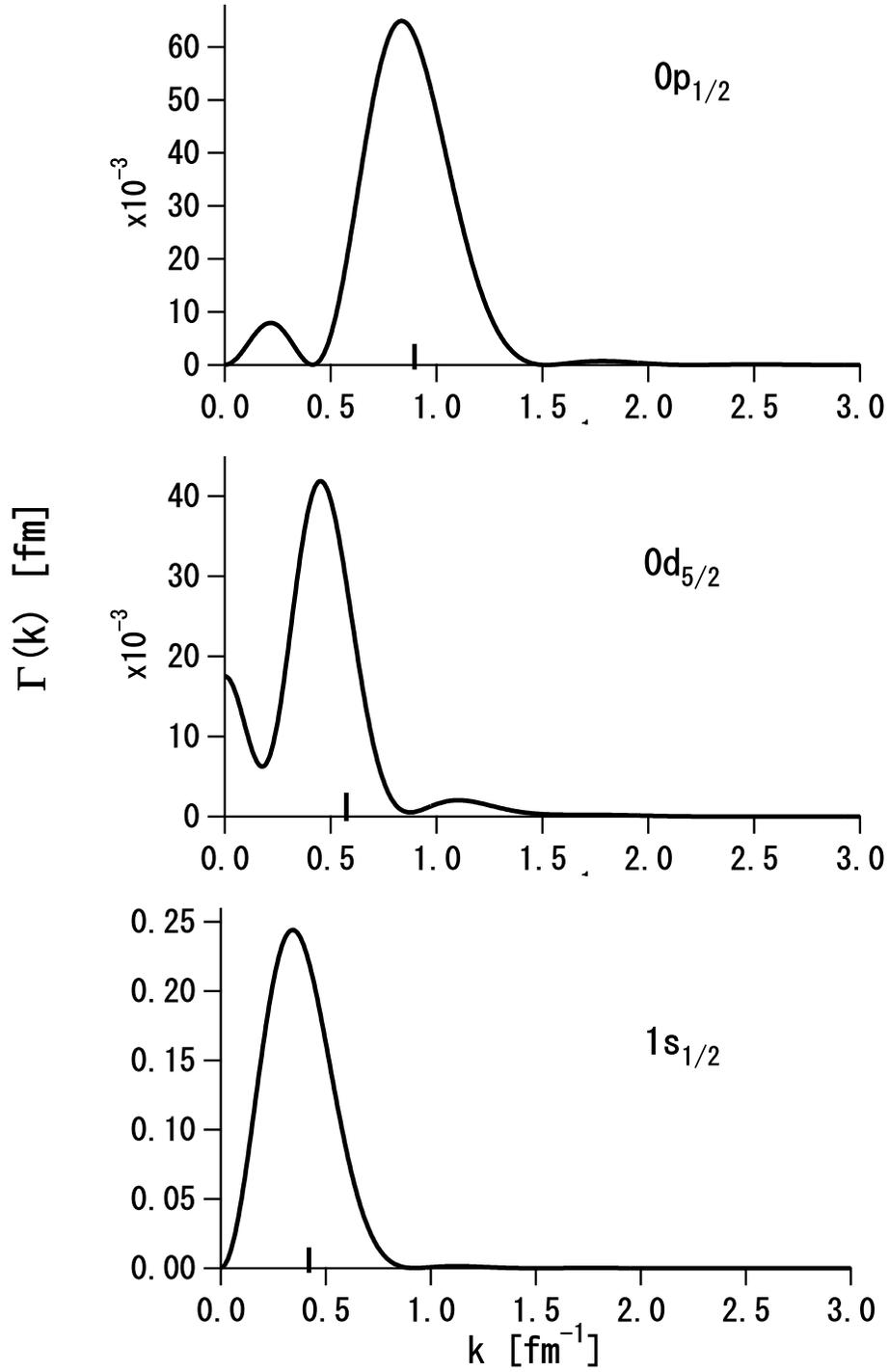}}
\vspace{2mm}
\caption{$\Gamma_{n\ell j}(k)$ of neutron q.p. levels in $^{26}$O
 calculated with the D1S interaction.
 The values of $p^{(\ell j)}_{n}$ for individual q.p. levels
 are presented by the ticks on the horizontal axes.
\label{fig:qpwf-k}}
\end{figure}

We next look into the wave functions of the q.p. states,
using Set~C.
Figure~\ref{fig:qpwf} shows the radial wave functions
of the neutron q.p. levels in $^{26}$O
which correspond to the $0s_{1/2}$, $1s_{1/2}$ and $2s_{1/2}$ orbits,
in terms of $|r\,V^{(\ell j)}_n(r)|^2$ and $|r\,U^{(\ell j)}_n(r)|^2$.
In this calculation $^{26}$O is so loosely bound
with $\lambda\simeq -0.4\,\mathrm{MeV}$
that all q.p. levels would lie in the continuum,
satisfying $\lambda+\epsilon_{n\ell j}>0$.
In this regard the q.p. levels shown in Fig.~\ref{fig:qpwf}
are considered to be resonance-like levels.
For all the three q.p. states, whose q.p. energies distribute
from $1.4\,\mathrm{MeV}$ to $40\,\mathrm{MeV}$,
the energy-dependent exponential asymptotics in $r\,V^{(\ell j)}_n(r)$
are represented appropriately by the superposition of the GEM bases.
We find oscillatory behavior in $r\,U^{(\ell j)}_n(r)$.
In order to view how appropriately
the oscillatory asymptotics are described,
we define the following quantity from the Fourier transform
of $r\,U^{(\ell j)}_n(r)$,
\begin{equation}
 \Gamma_{n\ell j}(k) \equiv \frac{1}{\pi}
  \left|\int_0^\infty r\,U^{(\ell j)}_n(r)\,e^{ikr}\,dr \right|^2\,
  \left(=\Gamma_{n\ell j}(-k)\right)\,.
\end{equation}
The scale of $\Gamma_{n\ell j}(k)$ is connected
to the occupation probability of the q.p. level, via
\begin{equation}
 \int_0^\infty \Gamma_{n\ell j}(k)\,dk =
  \int_0^\infty |U^{(\ell j)}_n(r)|^2\,r^2dr =
  1-\int_0^\infty |V^{(\ell j)}_n(r)|^2\,r^2dr\,.
\end{equation}
If $U^{(\ell j)}_n(r)$ fulfills the asymptotics of Eq.~(\ref{qpwf-asymp})
with $\lambda+\epsilon_{n\ell j}>0$,
$\Gamma_{n\ell j}(k)$ should have a peak at $k=p^{(\ell j)}_{n}$.
In Fig.~\ref{fig:qpwf-k} we show $\Gamma_{n\ell j}(k)$
for several low-lying q.p. levels of $^{26}$O.
Viewing the peaks of $\Gamma_{n\ell j}(k)$ at the positions
compatible with the q.p. energies,
we confirm that the energy-dependent oscillatory asymptotics
are properly expressed through the superposition of the GEM bases.

In Table~\ref{tab:rad_O}
we show the rms matter radii in HF and HFB
obtained through the three basis-sets.
Contribution of the center-of-mass motion is subtracted
as in Ref.~\cite{ref:Nak03}.
There is no significant difference among the bases up to $^{24}$O,
and this seems consistent with the results in Table~\ref{tab:eng_O}.
However, we find sizable basis-dependence in $^{25,26}$O,
although the energies do not differ significantly.
In $^{25}$O, where neutrons are not in the superfluid phase,
the s.p. energy of $n0d_{3/2}$ in the HF scheme is vanishing
($\approx -0.05\,\mathrm{MeV}$).
Thereby the calculated radius of $^{25}$O is extremely sensitive
to precise value of the s.p. energy.
The $^{26}$O case gives an illustration of the subtlety
for the drip-line nuclei that have no discrete q.p. levels,
as has been discussed in Sec.~\ref{sec:asymp}.
In these cases in which the nuclei are very close
to the neutron drip line,
some physical quantities (\textit{e.g.} nuclear radius) may depend
on details of the calculation.
Comparison of the radii between the HF and HFB calculations
will be of a certain interest.
The matter radii in the HFB scheme are similar to those in HF scheme
around the $\beta$-stability line.
However, in $^{22}$O the HFB calculations give slightly larger radius
than the HF calculations,
primarily due to the partial occupation of $n1s_{1/2}$.
Contrastingly, in $^{26}$O the HFB radius is substantially
smaller than the HF one in Sets~B and C.
This corresponds to the pairing anti-halo effect~\cite{ref:BDP00},
although attention should be paid to the subtlety
discussed in Sec.~\ref{sec:asymp}.

\subsection{Dependence of shell structure on HF interaction}

It has been argued that the spin-isospin nature
of the $NN$ interaction could play a role
in producing the new magic numbers~\cite{ref:Vst}.
We have developed an M3Y-type effective interaction
called M3Y-P2~\cite{ref:Nak03},
which is applicable to the self-consistent HF calculations.
Keeping the one-pion-exchange contribution
to the central part of the $NN$ interaction,
the M3Y-P2 interaction possesses reasonable spin-isospin nature,
as was checked via the Landau-Migdal parameter in the nuclear matter.
Within the self-consistent HF calculations,
shell gaps at $N=16$~\cite{ref:Nak03} and at $N=32$~\cite{ref:Nak04}
behave quite differently between M3Y-P2
and frequently used interactions such as D1S.
With the D1S interaction the shell gap
between $n1s_{1/2}$ and $n0d_{3/2}$ in the $N=16$ isotones
is almost constant, as $Z$ changes from $8$ to $16$.
On the contrary, the shell gap becomes narrower as $Z$ increases,
if we adopt the M3Y-P2 interaction.
Similar behavior has been found in the $N=32$ isotones.
It should be pointed out that
major difference lies around the $\beta$-stable region,
rather than in the region close to the drip-line.
We confirmed that the $Z$-dependence in the M3Y-P2 result is
primarily ascribed to the contribution of the one-pion-exchange part.

The pair correlation can break the closed core,
even in the spherical nuclei.
This core breaking is realized in the HFB picture,
unless the relevant shell gap is large enough.
To investigate effects of the pair correlation
in the region concerning the new magic numbers,
we carry out the HFB calculations using the GEM algorithm
for the $N=16$ and $32$ nuclei.
While the D1S interaction has moderate pairing properties,
the pairing character of the M3Y-P2 interaction is not realistic,
since it has not been tuned.
Here we always use the D1S interaction for the pairing interaction,
\textit{i.e.} to obtain the pair potential
in the HFB equation (\ref{HFBeq2}),
and examine dependence of the magic nature of $N=16$ and $32$
on the HF interactions, for which we take D1S and M3Y-P2.
The $\ell\leq 4$ ($\ell\leq 5$) truncation is made
for the $N=16$ ($N=32$) nuclei.
For the basis-set,
Set~C in the preceding subsection is employed, with $K=12$.

In Table~\ref{tab:Epair},
the neutron pair energies in the $N=16$ and $32$ nuclei
are compared between the two HF interactions.
The pair energy is defined as contribution of the pairing tensor
to the energy of the system.
In the $N=16$ isotones, the D1S interaction does not give rise to
the superfluidity for neutrons in $8\leq Z\leq 14$,
as long as the spherical symmetry is assumed.
In this regard, the D1S interaction provides a prediction
that the $N=16$ magic number holds
in the whole region of $8\leq Z\leq 14$.
On the contrary, in the case that we use the M3Y-P2 interaction
for the HF potential,
the neutron pair energy increases as $Z$ goes from $8$ to $14$,
\textit{i.e.} from the drip-line region to the $\beta$-stable region.
This behavior of the neutron pair energies,
which is a reflection of shell structure in the HF scheme,
is typically viewed in the difference between $^{24}$O and $^{30}$Si.
Similar trend is seen in the $N=32$ isotones.
While in $^{52}$Ca the pair energy is not large and is close to each other
between the two HF interactions,
we find sizable interaction-dependence in $^{60}$Ni,
due to the difference in the shell structure in the HF regime.

\begin{figure}
\centerline{\includegraphics[width=13cm]{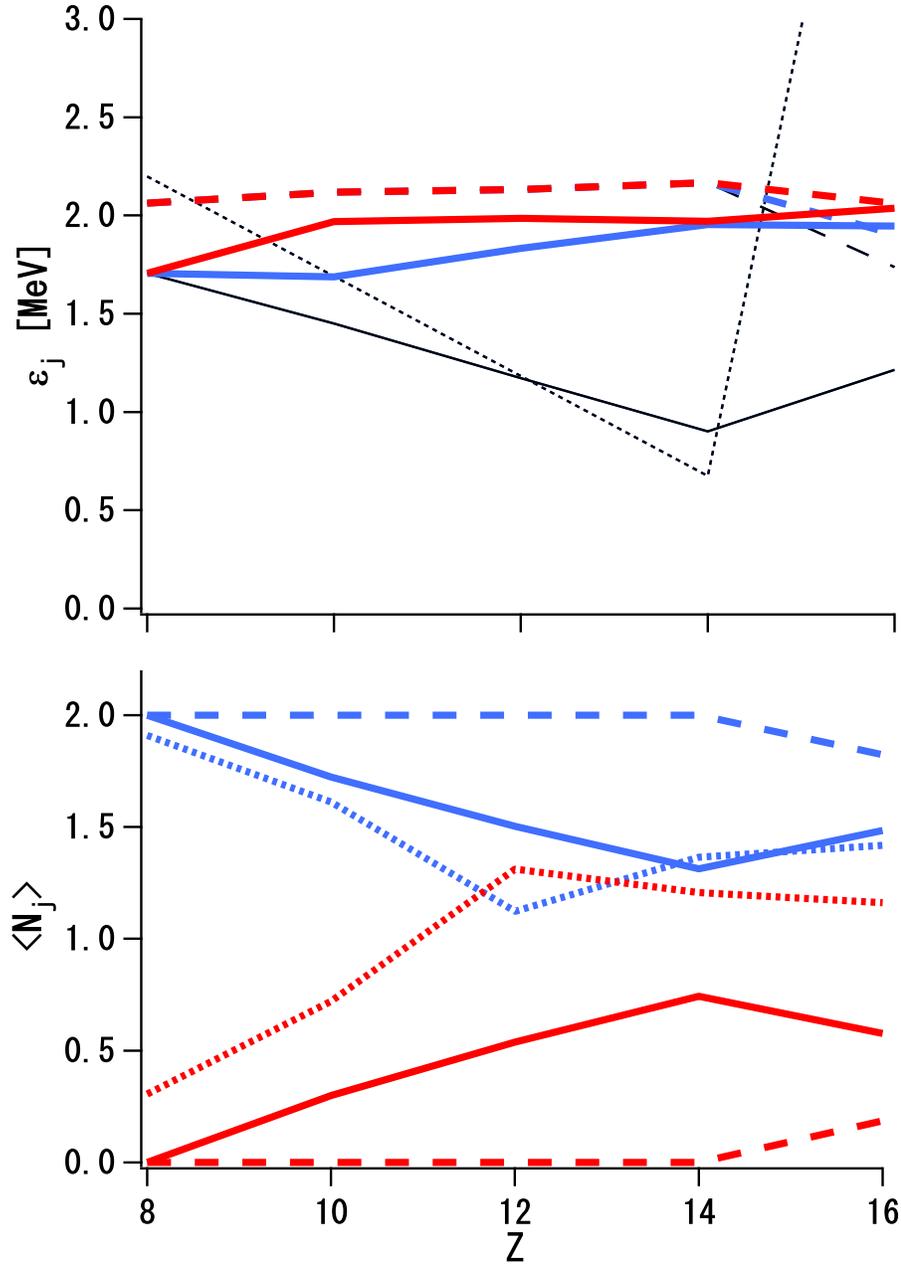}}
\vspace{2mm}
\caption{Neutron quasiparticle energies (upper panel)
 and neutron occupation numbers (lower panel)
 in the $Z=\mathrm{even}$, $N=16$ isotones.
 In both panels the red lines present the values of $0d_{3/2}$
 in the HFB calculations,
 while the blue lines those of $1s_{1/2}$ if visible.
 The full (dashed) lines with each color are obtained
 with the M3Y-P2 (the D1S) HF interaction.
 The thin black lines indicate the HF results
 (see text for details).
 The black dotted line in the upper panel
 is obtained from the HF treatment of the USD Hamiltonian.
 The dotted lines with each color in the lower panel
 represent the occupation numbers obtained
 from the full shell-model calculation with the USD Hamiltonian.
 \label{fig:N16qpe}}
\end{figure}

Effects of the pair correlation on excitation spectra
are basically viewed in the q.p. energies.
In the upper panel of Fig.~\ref{fig:N16qpe},
the neutron q.p. energies
corresponding to the $1s_{1/2}$ and $0d_{3/2}$ orbits are compared
between the D1S and the M3Y-P2 interactions
for the $Z=\mathrm{even}$, $N=16$ isotones.
For reference, we also plot the s.p. energies
measured from the Fermi energy in the HF approximation
without the sign,
as counterparts to the q.p. energies.
We here define the Fermi energy as the arithmetic average
of the HF energies of $n1s_{1/2}$ and $n0d_{3/2}$.
Hence, half the shell gap between $n1s_{1/2}$ and $n0d_{3/2}$
is presented as the HF results.
In the case that the neutron pair energy vanishes
in the HFB calculations,
we take the q.p. energies to be equal to their HF counterparts.
We also show the values obtained from a HF picture within the $sd$-shell,
employing the `USD' shell-model Hamiltonian~\cite{ref:USD}.
As pointed out in Ref.~\cite{ref:Vst},
the HF results indicate that the USD interaction
yields significant $Z$-dependence of the shell gap,
having qualitative similarity to the M3Y-P2 interaction.
It should be noticed that
the HF treatment of the shell-model Hamiltonian
tends to give radical variation of the s.p. energies
from nuclide to nuclide,
in comparison with the self-consistent mean-field approaches,
because effects of the rearrangement of the core are incorporated
into the interaction within the valence shell in an effective manner.

The HF and the HFB energies coincide up to $Z=14$
in the calculations using the D1S interaction,
because the nuclei are not in the superfluid phase.
In the HFB results, the M3Y-P2 interaction provides
almost constant q.p. energies of $n1s_{1/2}$ and $n0d_{3/2}$,
which are close to those given by the D1S interaction.
This is because the narrowing of the shell gap is compensated
by the pairing gap to a considerable extent.
It is suggested from the q.p. energies
that the two interactions yield similar excitation spectra.
In other words, the M3Y-P2 interaction does not seem to destroy
the energy spectra predicted in the widely used interactions
like D1S,
despite the difference in physical contents.
However, additional correlations could affect the energy levels
considerably.
It will be interesting to investigate
influence of the correlations beyond HFB in these nuclei.

The occupation numbers of the relevant q.p. levels are depicted
in the lower panel of Fig.~\ref{fig:N16qpe}.
Due to the narrowing shell gap,
we observe depletion of the $n1s_{1/2}$ occupation
in the HFB results with M3Y-P2,
as $Z$ grows from $8$ to $14$.
This indicates, consistently with the behavior of the pair energies,
that the magic nature of $N=16$ becomes weaker as $Z$ increases.
Thus the spectroscopic factors of the neutron orbits
could be useful in discriminating the two pictures;
a picture based on the almost constant $N=16$ shell gap,
and that on the decreasing shell gap for increasing $Z$
which may be compensated by the pairing gap.
As shown in Fig.~\ref{fig:N16qpe},
the occupation numbers obtained with M3Y-P2
are compared to those in the full shell-model calculation
using the USD Hamiltonian,
except at $Z=12$ where influence of the quadrupole deformation
seems important.

\begin{figure}
\centerline{\includegraphics[width=13cm]{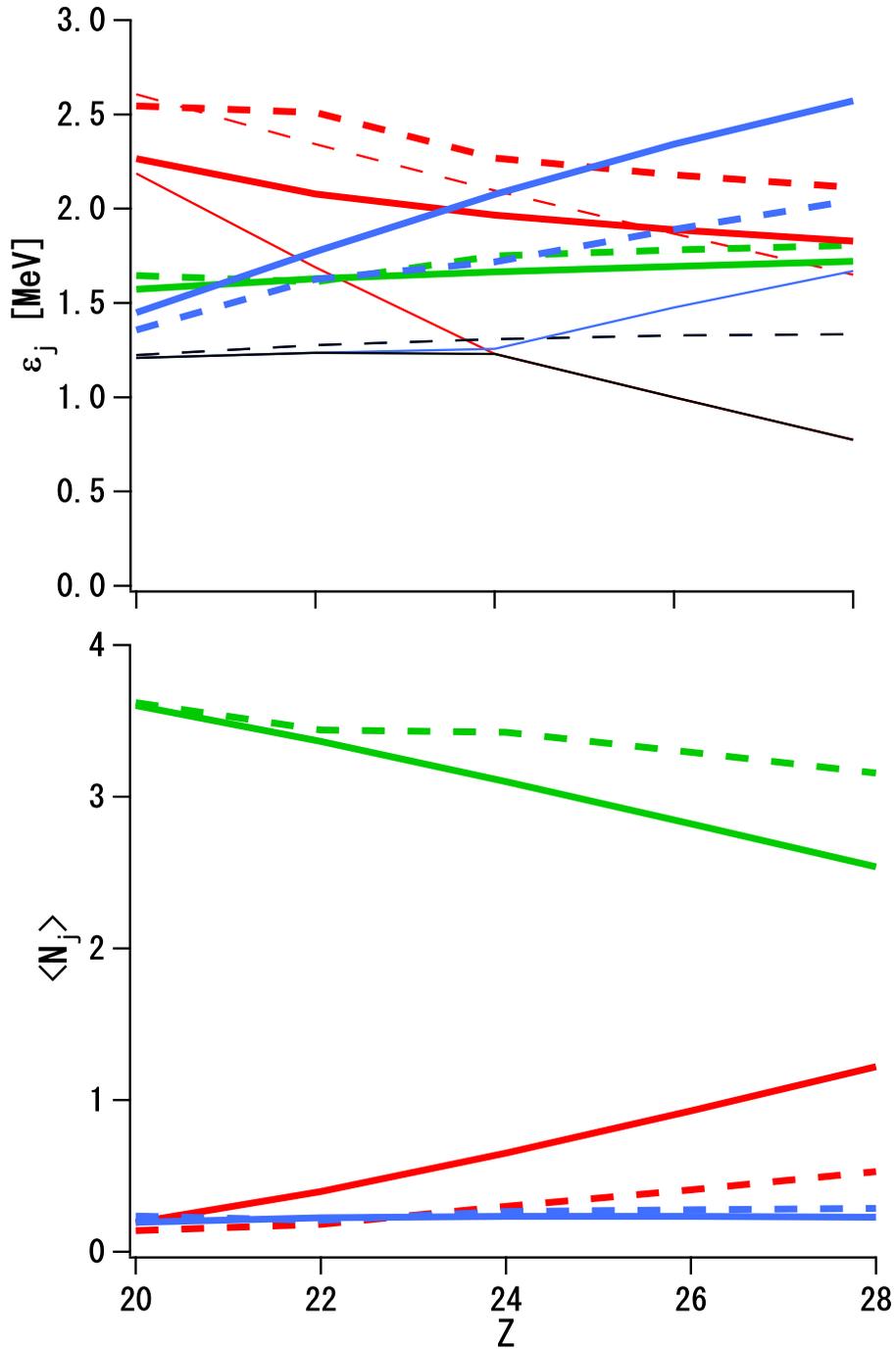}}
\vspace{2mm}
\caption{Neutron quasiparticle energies (upper panel)
 and neutron occupation numbers (lower panel)
 in the $Z=\mathrm{even}$, $N=32$ isotones.
 In both panels the red, blue and green lines present the values
 corresponding to the $0f_{5/2}$, $1p_{1/2}$ and $1p_{3/2}$ levels.
 The full (dashed) lines with each color are obtained
 with the M3Y-P2 (the D1S) HF interaction.
 The thick lines indicate the HFB results,
 while the thin lines the HF results.
 In the upper panel, the black color is used
 for the energy of the two levels adjacent to the Fermi energy
 in the HF results.
 \label{fig:N32qpe}}
\end{figure}

Figure~\ref{fig:N32qpe} shows the q.p. energies
and the occupation numbers of the levels
corresponding to the $n1p_{3/2}$, $n0f_{5/2}$ and $n1p_{1/2}$ orbits,
for the $Z=\mathrm{even}$, $N=32$ isotones.
As the HF counterparts to the q.p. energies,
we plot the s.p. energies measured from the Fermi energy,
without the sign.
The Fermi energy is taken to be the arithmetic average of the energies
of the two levels nearest the $N=32$ gap
(\textit{i.e.} $1p_{3/2}$
and the lower level out of $0f_{5/2}$ and $1p_{1/2}$).
As a result, the two levels have equal energy,
which represent half the shell gap,
in regard to the HF counterparts to the q.p. energies.
These values are shown by the thin black lines in the upper panel
of Fig.~\ref{fig:N32qpe}.
As argued in Ref.~\cite{ref:Nak04},
the D1S interaction gives almost constant shell gap in $20\leq Z\leq 28$,
while M3Y-P2 yields erosion of the gap
as $Z$ goes from $20$ to $28$.
Furthermore, the s.p. energy difference
between $n0f_{5/2}$ and $n1p_{3/2}$ changes more rapidly
in the M3Y-P2 results than in the D1S ones.
This gives rise to the inversion of $n0f_{5/2}$ and $n1p_{1/2}$
near $Z=24$ in the M3Y-P2 case,
while such inversion does not occur in the D1S case.
This shell structure in the HF scheme
affects the HFB occupation numbers.
Although the $N=32$ core does not close
in the whole region of $20\leq Z\leq 28$
in the strict sense of the HFB theory,
at $Z=20$ the occupation number of each orbit is not
significantly different from the value assuming the $N=32$ closure,
irrespectively of the HF potential.
If we use the D1S interaction for the HF potential,
this nature is almost preserved as $Z$ increases.
In contrast, when we apply the M3Y-P2 interaction,
leakage out of $n1p_{3/2}$ appreciably grows for increasing $Z$,
mainly due to the pair excitation to $n0f_{5/2}$.
Despite this difference in the physical picture,
the q.p. energies are much less interaction-dependent,
because the decreasing shell gap in M3Y-P2
is compensated by the increasing pairing gap to a certain extent.
Therefore the excitation spectra will not differ much
between the two HF interactions,
unless disturbance due to the additional correlations is
substantially strong.
Yet the inversion of $n0f_{5/2}$ and $n1p_{1/2}$
occurs in the HFB results with M3Y-P2,
which is not seen in the results with D1S.

\section{Summary}
\label{sec:summary}

Extending the previous work on the Hartree-Fock (HF) calculations
based on the Gaussian expansion method (GEM),
we develop a new method of implementing
the Hartree-Fock-Bogolyubov (HFB) calculations.
This method has two notable advantages:
(i) It is efficient in describing
the exponential and the oscillatory asymptotics
of wave functions at large $r$,
which depend on quasiparticle energy
and could play a significant role in nuclei close to the drip lines.
(ii) We can handle various effective interactions,
including those inducing non-local nuclear currents.

The present method has been tested numerically
in the oxygen isotopes with the Gogny D1S force,
by assuming the spherical symmetry and the parity conservation.
Three sets of GEM bases are tried and their results are compared;
a set comprised of the real-range Gaussians,
a set comprised of the complex-range Gaussians,
and a set of their mixture.
The last one gives the best results
in the context of the variational principle.
We have confirmed that both the exponential
and the oscillatory asymptotics
of the quasiparticle wave functions are properly described.

Relating to the new magic numbers in nuclei
far from the $\beta$-stability,
we have investigated interplay of the pair correlation
and the shell gaps at $N=16$ and $32$.
The Gogny D1S and the M3Y-P2 interactions are used
for the HF interaction,
while for the pair potential we always use the D1S interaction.
Reflecting the interaction-dependence of the shell structure
in the HF scheme,
the M3Y-P2 case provides stronger $Z$-dependence
of the superfluidity at $N=16$ and $32$ than the D1S case.
While the q.p. energies do not differ significantly
between the D1S and the M3Y-P2 interactions,
suggesting that excitation spectra could resemble each other,
the occupation numbers of neutron orbits
reflect the interaction-dependence of the shell structure.
The occupation numbers obtained from the M3Y-P2 interaction
are comparable with those in the standard shell model,
for the $N=16$ nuclei.
\\

\noindent
The author is grateful to Y. R. Shimizu for discussions.
This work is financially supported in part
as Grant-in-Aid for Scientific Research (B), No.~15340070,
by the Ministry of Education, Culture, Sports, Science and Technology,
Japan.
Numerical calculations are performed on HITAC SR11000
at Institute of Media and Information Technology, Chiba University,
and at Information Technology Center, University of Tokyo.

\appendix
\section*{Appendices}

\section{HFB equation for non-orthogonal bases}
\label{app:HFBeq}

In this appendix we present the HFB theory
in the case that non-orthogonal s.p. bases are adopted.
We denote the creation (annihilation) operator
associated with the s.p. basis $k$ by $c_k^\dagger$ ($c_k$).
The non-orthogonality leads to the non-canonical commutation relations,
\begin{equation}
 \{c_k, c_{k'}^\dagger\} = N_{kk'}\,,\quad
 \{c_k, c_{k'}\} = \{c_k^\dagger, c_{k'}^\dagger\} = 0\,,
\end{equation}
where $\mathsf{N}=(N_{kk'})$ is the norm matrix
satisfying $N_{kk'}^\ast=N_{k'k}$ and being positive-definite.
The generalized Bogolyubov transformation is defined by
\begin{equation}
 \alpha_i^\dagger = \sum_k \left(U_{ki}\,c_k^\dagger
			    + V_{ki}\,c_k\right)\,.
\label{Bog1}\end{equation}
In the matrix representation, Eq.~(\ref{Bog1}) can be expressed as
\begin{equation}
 (\begin{array}{cc} \mathsf{\alpha}^\dagger & \mathsf{\alpha}
  \end{array})
 = (\begin{array}{cc} \mathsf{c}^\dagger & \mathsf{c} \end{array})\,
 \mathsf{W}\,;\quad
 \mathsf{W} = \left(\begin{array}{cc} \mathsf{U} & \mathsf{V}^\ast \\
  \mathsf{V} & \mathsf{U}^\ast \end{array}\right)\,.
\label{Bog2}\end{equation}
Obtained as a solution of the HFB equation,
$\alpha_i^\dagger$ and $\alpha_i$ obey
the canonical commutation relations,
\begin{equation}
 \{\alpha_i, \alpha_{i'}^\dagger\} = \delta_{ii'}\,,\quad
 \{\alpha_i, \alpha_{i'}\} = \{\alpha_i^\dagger, \alpha_{i'}^\dagger\}
 = 0\,.
\end{equation}
This indicates
\begin{equation}
 \overline{\mathsf{W}}\,\mathsf{N}'\,\mathsf{W} = 1\,,\quad
  \mathsf{W}\,\overline{\mathsf{W}} = \mathsf{N}'^{-1}\,;\quad
 \mathsf{N}' = \left(\begin{array}{cc} \mathsf{N} & 0\\
     0 & \mathsf{N}^\ast\end{array}\right)\,,
\label{unitary1}\end{equation}
\textit{i.e.},
\begin{eqnarray}
 \overline{\mathsf{U}}\,\mathsf{N}\,\mathsf{U}
  + \overline{\mathsf{V}}\,\mathsf{N}^\ast\,\mathsf{V} = 1\,,&&
 \mathsf{V}^\mathrm{T}\,\mathsf{N}\,\mathsf{U}
  + \mathsf{U}^\mathrm{T}\,\mathsf{N}^\ast\,\mathsf{V} = 0\,,\nonumber\\
 \mathsf{U}\,\overline{\mathsf{U}} + \mathsf{V}^\ast\,\mathsf{V}^\mathrm{T}
  = \mathsf{N}^{-1}\,,&&
 \mathsf{U}\,\overline{\mathsf{V}} + \mathsf{V}^\ast\,\mathsf{U}^\mathrm{T}
  = 0\,.
\label{unitary2}\end{eqnarray}
We here use the expression $\overline{\mathsf{A}}=
(\mathsf{A}^\ast)^\mathrm{T}$, instead of $\mathsf{A}^\dagger$,
to avoid confusion with the hermitian conjugation of operators.
The inversion of the Bogolyubov transformation
of Eq.~(\ref{Bog2}) yields
\begin{equation}
 (\begin{array}{cc} \mathsf{c}^\dagger & \mathsf{c} \end{array})
 = (\begin{array}{cc} \mathsf{\alpha}^\dagger & \mathsf{\alpha}
    \end{array})\,
 \overline{\mathsf{W}}\,\mathsf{N}'\,.
\label{Bog3}\end{equation}

We define the density matrix and the pairing tensor by
\begin{eqnarray}
 (\mathsf{N}\,\mathsf{\rho}\,\mathsf{N})_{kk'}
  ~=~\langle\Phi|c_{k'}^\dagger c_k|\Phi\rangle
  &=& (\mathsf{N}\,\mathsf{V}^\ast\,\mathsf{V}^\mathrm{T}\,
  \mathsf{N})_{kk'}\,, \nonumber\\
 (\mathsf{N}\,\mathsf{\kappa}\,\mathsf{N}^\ast)_{kk'}
  ~=~\langle\Phi|c_{k'}c_k|\Phi\rangle
  &=& \langle\Phi|c_k^\dagger c_{k'}^\dagger|\Phi\rangle^\ast ~=~
  (\mathsf{N}\,\mathsf{V}^\ast\,\mathsf{U}^\mathrm{T}\,
  \mathsf{N}^\ast)_{kk'}\,,
\label{Rmat0}\end{eqnarray}
where $|\Phi\rangle$ is the HFB vacuum.
This HFB vacuum $|\Phi\rangle$ satisfies
$\alpha_{n\ell jm}|\Phi\rangle = 0$ and $\langle\Phi|\Phi\rangle = 1$.
Equation~(\ref{Rmat0}) obviously indicates
\begin{equation}
 \rho_{kk'} = (\mathsf{V}^\ast\,\mathsf{V}^\mathrm{T})_{kk'}\,,
 \quad \kappa_{kk'} = (\mathsf{V}^\ast\,\mathsf{U}^\mathrm{T})_{kk'}\,.
\label{Rmat}\end{equation}
In the HFB approximation the expectation value of the number operator
is adjusted to the actual particle number $n$.
This constraint is expressed
by $\mathrm{Tr}(\mathsf{N}\,\mathsf{\rho})=n$.
The HF Hamiltonian and the pair potential are defined by
\begin{eqnarray}
 h_{kk'} ~=~ \frac{\delta}{\delta\rho_{kk'}}\langle\Phi|\hat{H}|\Phi\rangle
  &=& \langle\Phi|\{[c_k, \hat{H}], c_{k'}^\dagger\}|\Phi\rangle\,,
  \nonumber\\
 \Delta_{kk'} ~=~ \frac{\delta}{\delta\kappa_{kk'}^\ast}\langle\Phi|
  \hat{H}|\Phi\rangle
  &=& \langle\Phi|\{[c_k, \hat{H}], c_{k'}\}|\Phi\rangle\,,
\label{spham0}\end{eqnarray}
from which the following HFB Hamiltonian matrix is obtained,
\begin{equation}
 \mathsf{\mathcal{H}} = \left(\begin{array}{cc}
  \mathsf{h}-\lambda & \mathsf{\Delta} \\
  -\mathsf{\Delta}^\ast & -\mathsf{h}^\ast+\lambda \end{array}\right)\,,
\end{equation}
with the chemical potential $\lambda$.
The HFB equation for the non-orthogonal bases has
the following form of the generalized eigenvalue equation,
\begin{equation}
 \mathsf{\mathcal{H}}\,\mathsf{W} = \mathsf{N}'\,\mathsf{W}
  \left(\begin{array}{cc} \mathsf{\epsilon} & 0\\
  0 & -\mathsf{\epsilon} \end{array}\right)\,;\quad
  \mathsf{\epsilon} = \mathrm{diag}(\epsilon_i)\,.
\end{equation}

It is commented that,
because $\mathsf{N}$ is a Hermite positive-definite matrix,
it can be decomposed as $\mathsf{N}=\mathsf{L}\,\overline{\mathsf{L}}$
by using a lower-triangular matrix $\mathsf{L}$.
Multiplication of $\mathsf{c}$ by $\mathsf{L}^{-1}$
gives an orthogonalization of the s.p. bases,
and the canonical commutation relation follows;
\begin{equation}
 \{(\mathsf{L}^{-1}\,\mathsf{c})_k,
  ((\mathsf{L}^{-1})^\ast\,\mathsf{c}^\dagger)_{k'}\}
  = \delta_{kk'}\,.
\end{equation}
All the equations return to those in the usual HFB theory
for the orthogonal bases,
by multiplying by $\mathsf{L}^{-1}$ appropriately.

For the Hamiltonian consisting of the one-body term $\hat{K}$
and the two-body interaction $\hat{V}$,
we obtain
\begin{equation}
 \langle\Phi|\hat{H}|\Phi\rangle
  = \sum_{k_1 k_2} \langle k_2|\hat{K}|k_1\rangle\,\rho_{k_1 k_2}
  + \frac{1}{4}\sum_{k_1 k_2 k_3 k_4}
  \langle k_3 k_4|\hat{V}|k_1 k_2\rangle\,
  (2\rho_{k_1 k_3}\rho_{k_2 k_4} + \kappa_{k_1 k_2}\kappa^\ast_{k_3 k_4})
  \,,
\label{HFB-E2}\end{equation}
where $|k\rangle=c_k^\dagger|0\rangle$ and
$|kk'\rangle=c_k^\dagger c_{k'}^\dagger|0\rangle$,
with $|0\rangle$ representing the particle vacuum
satisfying $c_k|0\rangle=0$.
Equation~(\ref{HFB-E2}) derives
\begin{eqnarray}
 h_{kk'} &=& \langle k'|\hat{K}|k\rangle
  + \sum_{k_1 k_2} \langle k' k_2|\hat{V}|k k_1\rangle\,\rho_{k_1 k_2}\,,
  \nonumber\\
 \Delta_{kk'} &=& \frac{1}{2}\sum_{k_1 k_2}
  \langle k k'|\hat{V}|k_1 k_2\rangle\,\kappa_{k_1 k_2}\,.
\end{eqnarray}
For interactions having the form of Eq.~(\ref{eq:effint}),
rearrangement terms for $v_{12}^{(\mathrm{DD})}$,
which emerge via variation of $\rho^\alpha$,
should be taken into account.

\section{HFB equation with spherical symmetry}
\label{app:sphHFB}

By adopting the spherically symmetric s.p. bases
$k=(\nu\ell jm)$ and $k'=(\nu'\ell'j'm')$,
the norm matrix $\mathsf{N}$ is simplified as in Eq.~(\ref{eq:norm}),
\begin{equation}
 N_{kk'} = \delta_{\ell\ell'}\delta_{jj'}\delta_{mm'}\cdot
  N^{(\ell j)}_{\nu\nu'}\,.
\end{equation}
If we assume the spherical symmetry on the HFB fields,
the q.p. state is taken to be $i=(n\ell jm)$,
and the $U$- and $V$-coefficients in Eq.~(\ref{Bog1}) have the form
\begin{equation}
 U_{k'i} = \delta_{\ell\ell'}\delta_{jj'}\delta_{mm'}\cdot
  U^{(\ell j)}_{\nu'n}\,,\quad
 V_{k'i} = \delta_{\ell\ell'}\delta_{jj'}\delta_{m-m'}(-)^{j+m}\cdot
  V^{(\ell j)}_{\nu'n}\,.
\label{sphUV}\end{equation}
Equation~(\ref{unitary2}) is then reduced to
\begin{eqnarray}
 \overline{\mathsf{U}^{(\ell j)}}\,\mathsf{N}^{(\ell j)}\,
  \mathsf{U}^{(\ell j)}
  + \overline{\mathsf{V}^{(\ell j)}}\,\mathsf{N}^{(\ell j)\ast}\,
  \mathsf{V}^{(\ell j)} = 1\,,&&
 \mathsf{V}^{(\ell j)\mathrm{T}}\,\mathsf{N}^{(\ell j)}\,
  \mathsf{U}^{(\ell j)}
  - \mathsf{U}^{(\ell j)\mathrm{T}}\,\mathsf{N}^{(\ell j)\ast}\,
  \mathsf{V}^{(\ell j)} = 0\,,\nonumber\\
 \mathsf{U}^{(\ell j)}\,\overline{\mathsf{U}^{(\ell j)}}
  + \mathsf{V}^{(\ell j)\ast}\,\mathsf{V}^{(\ell j)\mathrm{T}}
  = \mathsf{N}^{(\ell j)-1}\,,&&
 \mathsf{U}^{(\ell j)}\,\overline{\mathsf{V}^{(\ell j)}}
 - \mathsf{V}^{(\ell j)\ast}\,\mathsf{U}^{(\ell j)\mathrm{T}}
  = 0\,.
\label{unitary3}\end{eqnarray}
Analogously to Eq.~(\ref{sphUV}),
the density matrix, the pairing tensor, the HF Hamiltonian
and the pair potential are taken as
\begin{eqnarray}
 \rho_{kk'}=\delta_{\ell\ell'}\delta_{jj'}\delta_{mm'}\cdot
  \rho^{(\ell j)}_{\nu\nu'}\,;&&
 \rho^{(\ell j)}_{\nu\nu'}=
  (\mathsf{V}^{(\ell j)\ast}\,\mathsf{V}^{(\ell j)\mathrm{T}})_{\nu\nu'}
  \,,\nonumber\\
 \kappa_{kk'}=\delta_{\ell\ell'}\delta_{jj'}\delta_{m-m'}(-)^{j-m}\cdot
  \kappa^{(\ell j)}_{\nu\nu'}\,;&&
 \kappa^{(\ell j)}_{\nu\nu'}=
  (\mathsf{V}^{(\ell j)\ast}\,\mathsf{U}^{(\ell j)\mathrm{T}})_{\nu\nu'}
  \,,
\label{Rmat2}\end{eqnarray}
and
\begin{equation}
 h_{kk'}=\delta_{\ell\ell'}\delta_{jj'}\delta_{mm'}\cdot
  h^{(\ell j)}_{\nu\nu'}\,,\quad
 \Delta_{kk'}=\delta_{\ell\ell'}\delta_{jj'}\delta_{m-m'}(-)^{j-m}\cdot
  \Delta^{(\ell j)}_{\nu\nu'}\,.
\label{spham}\end{equation}
Note that $\kappa^{(\ell j)}_{\nu\nu'}=\kappa^{(\ell j)}_{\nu'\nu}$
and $\Delta^{(\ell j)}_{\nu\nu'}=\Delta^{(\ell j)}_{\nu'\nu}$.
With these replacements the HFB equation under the spherical symmetry
is written as
\begin{equation}
 \mathsf{\mathcal{H}}^{(\ell j)}\,\mathsf{W}^{(\ell j)} =
  \mathsf{N}'^{(\ell j)}\,\mathsf{W}^{(\ell j)}
  \left(\begin{array}{cc} \mathsf{\epsilon}^{(\ell j)} & 0\\
  0 & -\mathsf{\epsilon}^{(\ell j)} \end{array}\right)\,;\quad
  \mathsf{\epsilon}^{(\ell j)} = \mathrm{diag}(\epsilon_{n\ell j})\,,
\label{HFBeq2}\end{equation}
where
\begin{eqnarray}
 \mathsf{\mathcal{H}}^{(\ell j)} = \left(\begin{array}{cc}
  \mathsf{h}^{(\ell j)}-\lambda & \mathsf{\Delta}^{(\ell j)} \\
  \mathsf{\Delta}^{(\ell j)\ast} & -\mathsf{h}^{(\ell j)\ast}+\lambda
 \end{array}\right)\,,\quad
 \mathsf{W}^{(\ell j)} = \left(\begin{array}{cc}
  \mathsf{U}^{(\ell j)} & \mathsf{V}^{(\ell j)\ast} \\
  -\mathsf{V}^{(\ell j)} & \mathsf{U}^{(\ell j)\ast} \end{array}\right)
 \,,\nonumber\\
 \mathsf{N}'^{(\ell j)} = \left(\begin{array}{cc}
   \mathsf{N}^{(\ell j)} & 0\\
   0 & \mathsf{N}^{(\ell j)\ast}\end{array}\right)\,.
\end{eqnarray}

\clearpage\thispagestyle{empty}
\begin{table}
\begin{center}
\caption{HF and HFB energies (MeV) of $^{14-26}$O,
 calculated with the D1S interaction.
\label{tab:eng_O}}
\begin{tabular}{crrrcrrr}
\hline\hline
nuclide & \multicolumn{3}{c}{HF} &~~& \multicolumn{3}{c}{HFB} \\
& \multicolumn{1}{c}{Set A} & \multicolumn{1}{c}{Set B}
& \multicolumn{1}{c}{Set C} && \multicolumn{1}{c}{Set A}
& \multicolumn{1}{c}{Set B} & \multicolumn{1}{c}{Set C} \\ \hline
$^{14}$O & $-~99.522$ & $-~99.533$ & $-~99.559$
&& \multicolumn{1}{c}{---} & $-~99.586$ & $-~99.613$\\
$^{15}$O & $-114.467$ & $-114.479$ & $-114.509$
&& $-114.467$ & $-114.479$ & $-114.509$\\
$^{16}$O & $-129.477$ & $-129.483$ & $-129.515$
&& $-129.477$ & $-129.483$ & $-129.515$\\
$^{17}$O & $-134.518$ & $-134.521$ & $-134.548$
&& $-134.518$ & $-134.521$ & $-134.548$\\
$^{18}$O & $-139.711$ & $-139.710$ & $-139.734$
&& $-142.283$ & $-142.286$ & $-142.309$\\
$^{19}$O & $-145.079$ & $-145.076$ & $-145.096$
&& $-146.391$ & $-146.391$ & $-146.411$\\
$^{20}$O & $-150.644$ & $-150.639$ & $-150.657$
&& $-153.329$ & $-153.331$ & $-153.350$\\
$^{21}$O & $-156.425$ & $-156.419$ & $-156.434$
&& $-156.425$ & $-156.419$ & $-156.434$\\
$^{22}$O & $-162.438$ & $-162.432$ & $-162.444$
&& $-162.540$ & $-162.546$ & $-162.563$\\
$^{23}$O & $-165.278$ & $-165.298$ & $-165.321$
&& $-165.278$ & $-165.298$ & $-165.321$\\
$^{24}$O & $-168.540$ & $-168.568$ & $-168.595$
&& $-168.540$ & $-168.568$ & $-168.595$\\
$^{25}$O & $-168.082$ & $-168.084$ & $-168.134$
&& $-168.082$ & $-168.084$ & $-168.134$\\
$^{26}$O & $-167.942$ & $-167.935$ & $-167.995$
&& $-169.248$ & $-169.266$ & $-169.304$\\
\hline\hline
\end{tabular}
\end{center}
\end{table}

\clearpage\thispagestyle{empty}
\begin{table}
\begin{center}
\caption{Rms radii (fm) of $^{14-26}$O in the HF and HFB calculations
 with the D1S interaction.
\label{tab:rad_O}}
\begin{tabular}{crrrcrrr}
\hline\hline
nuclide & \multicolumn{3}{c}{HF} &~~& \multicolumn{3}{c}{HFB} \\
& \multicolumn{1}{c}{Set A} & \multicolumn{1}{c}{Set B}
& \multicolumn{1}{c}{Set C} && \multicolumn{1}{c}{Set A}
& \multicolumn{1}{c}{Set B} & \multicolumn{1}{c}{Set C} \\ \hline
$^{14}$O & $2.532$ & $2.530$ & $2.530$
&& \multicolumn{1}{c}{---} & $2.535$ & $2.534$\\
$^{15}$O & $2.570$ & $2.568$ & $2.568$
&& $2.570$ & $2.568$ & $2.568$\\
$^{16}$O & $2.609$ & $2.607$ & $2.606$
&& $2.609$ & $2.607$ & $2.606$\\
$^{17}$O & $2.663$ & $2.662$ & $2.661$
&& $2.663$ & $2.662$ & $2.661$\\
$^{18}$O & $2.712$ & $2.711$ & $2.711$
&& $2.707$ & $2.706$ & $2.705$\\
$^{19}$O & $2.756$ & $2.755$ & $2.755$
&& $2.755$ & $2.754$ & $2.753$\\
$^{20}$O & $2.796$ & $2.795$ & $2.795$
&& $2.794$ & $2.793$ & $2.793$\\
$^{21}$O & $2.832$ & $2.831$ & $2.831$
&& $2.832$ & $2.831$ & $2.831$\\
$^{22}$O & $2.864$ & $2.863$ & $2.863$
&& $2.875$ & $2.874$ & $2.874$\\
$^{23}$O & $2.950$ & $2.947$ & $2.948$
&& $2.950$ & $2.947$ & $2.948$\\
$^{24}$O & $3.016$ & $3.013$ & $3.014$
&& $3.016$ & $3.013$ & $3.014$\\
$^{25}$O & $3.133$ & $3.102$ & $3.119$
&& $3.133$ & $3.102$ & $3.119$\\
$^{26}$O & $3.205$ & $3.177$ & $3.197$
&& $3.203$ & $3.159$ & $3.164$\\
\hline\hline
\end{tabular}
\end{center}
\end{table}

\begin{table}
\begin{center}
\caption{Interaction-dependence of neutron pair energies (MeV)
 in $N=16$ and $32$ nuclei.
\label{tab:Epair}}
\begin{tabular}{crr}
\hline\hline
nuclide & \multicolumn{1}{c}{D1S} & \multicolumn{1}{c}{M3Y-P2} \\ \hline
$^{24}$O & $0.000$ & $0.000$ \\
$^{30}$Si & $0.000$ & $-3.911$ \\ \hline
$^{52}$Ca & $-2.122$ & $-2.224$ \\
$^{60}$Ni & $-4.825$ & $-7.055$ \\
\hline\hline
\end{tabular}
\end{center}
\end{table}

\clearpage\thispagestyle{empty}

\clearpage\thispagestyle{empty}

\clearpage\thispagestyle{empty}

\clearpage\thispagestyle{empty}

\end{document}